# Quasi-periodic non-stationary solutions of 3D Euler equations for incompressible flow

**Sergey V. Ershkov**,

Institute for Time Nature Explorations,

M.V. Lomonosov's Moscow State University,

Leninskie gory, 1-12, Moscow 119991, Russia

e-mail: sergej-ershkov@yandex.ru

A novel derivation of non-stationary solutions of 3D Euler equations for incompressible inviscid flow is considered here. Such a solution is the product of 2 separated parts: - one consisting of the spatial component and the other being related to the time dependent part.

Spatial part of a solution could be determined if we substitute such a solution to the equations of motion (equation of momentum) with the requirement of *scale-similarity* in regard to the proper component of spatial velocity. So, the time-dependent part of equations of momentum should depend on the time-parameter only.

The main result, which should be outlined, is that the governing (time-dependent) ODE-system consist of 2 *Riccati*-type equations in regard to each other, which has no solution in general case. But we obtain conditions when each component of time-dependent part is proved to be determined by the proper *elliptical* integral in regard to the time-parameter *t*, which is a generalization of the class of inverse periodic functions.

## 1. Introduction: the Euler system of equations.

In accordance with [1-3], the Euler system of equations for incompressible flow of inviscid fluid should be presented in the Cartesian coordinates as below (under the proper initial conditions):

$$\nabla\cdot\vec{u} = 0\,, \tag{1.1}$$

$$\frac{\partial\vec{u}}{\partial t} + (\vec{u}\cdot\nabla)\vec{u} = -\frac{\nabla p}{\rho} + \vec{F}\,, \tag{1.2}$$

- where $\boldsymbol{u}$ is the flow velocity, a vector field; $\rho$ is the fluid density, $p$ is the pressure, $\boldsymbol{F}$ represents external force (*per unit of mass in a volume*) acting on the fluid; besides, we assume external force $\boldsymbol{F}$ above to be the *central* force, which has a potential $\phi$ represented by $\boldsymbol{F} = -\nabla\phi$ .

## 2. The originating system of PDE for Euler Eqs.

Using the identity $(\boldsymbol{u}\cdot\nabla)\boldsymbol{u} = (1/2)\nabla(\boldsymbol{u}^2) - \boldsymbol{u}\times(\nabla\times\boldsymbol{u})$, we could present the Euler equations in the case of incompressible flow of inviscid fluid $\boldsymbol{u} = \{u_1, u_2, u_3\}$ as below [4-5]:

$$\nabla\cdot\vec{u} = 0\,,$$
$$\frac{\partial\vec{u}}{\partial t} = \vec{u}\times\vec{\omega} - \left(\frac{1}{2}\nabla(\vec{u}^{\,2}) + \frac{\nabla p}{\rho} + \nabla\phi\right) \tag{2.1}$$

- here we denote *the curl field* $\boldsymbol{\omega} = \nabla\times\boldsymbol{u}$, a pseudovector field (*time-dependent*) [6]:

$$\{\omega_x, \omega_y, \omega_z\} \equiv \left\{\left(\frac{\partial u_3}{\partial y} - \frac{\partial u_2}{\partial z}\right),\ \left(\frac{\partial u_1}{\partial z} - \frac{\partial u_3}{\partial x}\right),\ \left(\frac{\partial u_2}{\partial x} - \frac{\partial u_1}{\partial y}\right)\right\} \tag{2.2}$$

- also we denote $\nabla\phi = \{f_x, f_y, f_z\}$ in (2.1); besides, let us choose $\rho = 1$ for simplicity.

## 3. Conditions for the space-part of exact solution.

Let us search for solutions $\{\boldsymbol{u}, p\}$ of the system (2.1) in a form below:

$$u_1 = U(t)\cdot u(x,y,z),\;\; u_2 = V(t)\cdot v(x,y,z),\;\; u_3 = W(t)\cdot w(x,y,z),\;\; p = P(t)\cdot p(x,y,z) \qquad (3.1)$$

- then we should obtain from (2.1) and expression (2.2) the proper system of PDE:

$$\begin{cases} \dfrac{\partial u_1}{\partial t} = (u_2\cdot\omega_z - u_3\cdot\omega_y) - \dfrac{1}{2}\dfrac{\partial}{\partial x}(u_1^{\,2} + u_2^{\,2} + u_3^{\,2}) - \dfrac{\partial}{\partial x}p - f_x\,, \\[2ex] \dfrac{\partial u_2}{\partial t} = (u_3\cdot\omega_x - u_1\cdot\omega_z) - \dfrac{1}{2}\dfrac{\partial}{\partial y}(u_1^{\,2} + u_2^{\,2} + u_3^{\,2}) - \dfrac{\partial}{\partial y}p - f_y\,, \\[2ex] \dfrac{\partial u_3}{\partial t} = (u_1\cdot\omega_y - u_2\cdot\omega_x) - \dfrac{1}{2}\dfrac{\partial}{\partial z}(u_1^{\,2} + u_2^{\,2} + u_3^{\,2}) - \dfrac{\partial}{\partial z}p - f_z\,, \end{cases} \qquad (3.2)$$

Besides, there exists the proper restriction from continuity equation (1.1) as below ($\partial u/\partial x \neq 0$):

$$U(t)\frac{\partial u}{\partial x} + V(t)\frac{\partial v}{\partial y} + W(t)\frac{\partial w}{\partial z} = 0\,, \;\; \Rightarrow$$

$$\frac{\partial v}{\partial y} = \chi\frac{\partial u}{\partial x}, \quad \frac{\partial w}{\partial z} = \lambda\frac{\partial u}{\partial x}, \quad \{\chi,\lambda\} = const \qquad (3.3)$$

The system of equations (3.2) should be transformed under conditions (3.1) as below:

$$
\begin{cases}
\dfrac{dU(t)}{dt} = \dfrac{V(t)\cdot v(x,y,z)\cdot\left(V(t)\cdot\dfrac{\partial v}{\partial x} - U(t)\cdot\dfrac{\partial u}{\partial y}\right) - W(t)\cdot w(x,y,z)\cdot\left(U(t)\cdot\dfrac{\partial u}{\partial z} - W(t)\cdot\dfrac{\partial w}{\partial x}\right)}{u(x,y,z)} - \\
- \dfrac{1}{2}\dfrac{\dfrac{\partial}{\partial x}(U(t)^2\cdot u^2(x,y,z) + V^2(t)\cdot v^2(x,y,z) + W^2(t)\cdot w^2(x,y,z))}{u(x,y,z)} - \dfrac{\left(P(t)\cdot\dfrac{\partial}{\partial x}p(x,y,z) + f_x\right)}{u(x,y,z)}, \\
\dfrac{dV(t)}{dt} = \dfrac{W(t)\cdot w(x,y,z)\cdot\left(W(t)\cdot\dfrac{\partial w}{\partial y} - V(t)\cdot\dfrac{\partial v}{\partial z}\right) - U(t)\cdot u(x,y,z)\cdot\left(V(t)\cdot\dfrac{\partial v}{\partial x} - U(t)\cdot\dfrac{\partial u}{\partial y}\right)}{v(x,y,z)} - \\
- \dfrac{1}{2}\dfrac{\dfrac{\partial}{\partial y}(U(t)^2\cdot u^2(x,y,z) + V^2(t)\cdot v^2(x,y,z) + W^2(t)\cdot w^2(x,y,z))}{v(x,y,z)} - \dfrac{\left(P(t)\cdot\dfrac{\partial}{\partial y}p(x,y,z) + f_y\right)}{v(x,y,z)}, \\
\dfrac{dW(t)}{dt} = \dfrac{U(t)\cdot u(x,y,z)\cdot\left(U(t)\cdot\dfrac{\partial u}{\partial z} - W(t)\cdot\dfrac{\partial w}{\partial x}\right) - V(t)\cdot v(x,y,z)\cdot\left(W(t)\cdot\dfrac{\partial w}{\partial y} - V(t)\cdot\dfrac{\partial v}{\partial z}\right)}{w(x,y,z)} - \\
- \dfrac{1}{2}\dfrac{\dfrac{\partial}{\partial z}(U(t)^2\cdot u^2(x,y,z) + V^2(t)\cdot v^2(x,y,z) + W^2(t)\cdot w^2(x,y,z))}{w(x,y,z)} - \dfrac{\left(P(t)\cdot\dfrac{\partial}{\partial z}p(x,y,z) + f_z\right)}{w(x,y,z)},
\end{cases}
\tag{3.4}
$$

- thus, from the 1-st of Eqs. (3.4) we should assume ($\{a_i\} = const$, $i = 1, \dots, 9$):

$$
\frac{v(x,y,z)\cdot\dfrac{\partial v}{\partial x}}{u(x,y,z)} = a_1, \quad -\frac{v(x,y,z)\cdot\dfrac{\partial u}{\partial y}}{u(x,y,z)} = a_2, \quad -\frac{w(x,y,z)\cdot\dfrac{\partial u}{\partial z}}{u(x,y,z)} = a_3, \quad \frac{w(x,y,z)\cdot\dfrac{\partial w}{\partial x}}{u(x,y,z)} = a_4,
$$

$$
-\frac{1}{2}\frac{\dfrac{\partial}{\partial x}(u^2(x,y,z))}{u(x,y,z)} = a_5, \quad -\frac{1}{2}\frac{\dfrac{\partial}{\partial x}(v^2(x,y,z))}{u(x,y,z)} = a_6, \quad -\frac{1}{2}\frac{\dfrac{\partial}{\partial x}(w^2(x,y,z))}{u(x,y,z)} = a_7, \tag{3.5}
$$

$$
-\frac{\dfrac{\partial}{\partial x}p(x,y,z)}{u(x,y,z)} = a_8, \quad -\frac{f_x}{u(x,y,z)} = a_9,
$$

- but the 2-nd of Eqs. (3.4) yields as below ($\{b_i\}$ = *const*, *i =1,…,9*):

$$\frac{w(x,y,z)\cdot\frac{\partial w}{\partial y}}{v(x,y,z)}=b_1,\quad -\frac{w(x,y,z)\cdot\frac{\partial v}{\partial z}}{v(x,y,z)}=b_2,\quad -\frac{u(x,y,z)\cdot\frac{\partial v}{\partial x}}{v(x,y,z)}=b_3,\quad \frac{u(x,y,z)\cdot\frac{\partial u}{\partial y}}{v(x,y,z)}=b_4,$$

$$-\frac{1}{2}\frac{\frac{\partial}{\partial y}(u^2(x,y,z))}{v(x,y,z)}=b_5,\quad -\frac{1}{2}\frac{\frac{\partial}{\partial y}(v^2(x,y,z))}{v(x,y,z)}=b_6,\quad -\frac{1}{2}\frac{\frac{\partial}{\partial y}(w^2(x,y,z))}{v(x,y,z)}=b_7, \qquad (3.6)$$

$$-\frac{\frac{\partial}{\partial y}p(x,y,z)}{v(x,y,z)}=b_8,\quad -\frac{f_y}{v(x,y,z)}=b_9,$$

- besides, 3-d of Eqs. (3.4) yields ($\{c_i\}$ = *const*, *i =1,…,9*):

$$\frac{u(x,y,z)\cdot\frac{\partial u}{\partial z}}{w(x,y,z)}=c_1,\quad -\frac{u(x,y,z)\cdot\frac{\partial w}{\partial x}}{w(x,y,z)}=c_2,\quad -\frac{v(x,y,z)\cdot\frac{\partial w}{\partial y}}{w(x,y,z)}=c_3,\quad \frac{v(x,y,z)\cdot\frac{\partial v}{\partial z}}{w(x,y,z)}=c_4,$$

$$-\frac{1}{2}\frac{\frac{\partial}{\partial z}(u^2(x,y,z))}{w(x,y,z)}=c_5,\quad -\frac{1}{2}\frac{\frac{\partial}{\partial z}(v^2(x,y,z))}{w(x,y,z)}=c_6,\quad -\frac{1}{2}\frac{\frac{\partial}{\partial z}(w^2(x,y,z))}{w(x,y,z)}=c_7, \qquad (3.7)$$

$$-\frac{\frac{\partial}{\partial z}p(x,y,z)}{w(x,y,z)}=c_8,\quad -\frac{f_z}{w(x,y,z)}=c_9\,.$$

## 4. <u>The space-part of exact solution.</u>

As for the structure of space part of exact solution (3.1), the system of equations (3.5)-(3.7) could be solved by the proper analytical way as below:

- Eq. (3.5) yields

$$\frac{v(x,y,z)\cdot\dfrac{\partial v}{\partial x}}{u(x,y,z)}=a_1,\quad \frac{\partial u}{\partial y}=-\left(\frac{a_2}{a_1}\right)\cdot\frac{\partial v}{\partial x},\quad \frac{w(x,y,z)\cdot\dfrac{\partial u}{\partial z}}{u(x,y,z)}=-a_3,\quad \frac{\partial w}{\partial x}=-\left(\frac{a_4}{a_3}\right)\cdot\frac{\partial u}{\partial z},$$

$$\frac{\partial}{\partial x}(u(x,y,z))=-a_5\quad a_6=-a_1,\quad a_7=-a_4,\quad \frac{\partial}{\partial x}p(x,y,z)=a_8\cdot u(x,y,z),\quad f_x=-a_9\cdot u(x,y,z), \tag{4.1}$$

- Eq. (3.6) yields

$$\frac{w(x,y,z)\cdot\dfrac{\partial w}{\partial y}}{v(x,y,z)}=b_1,\quad \frac{\partial w}{\partial y}=-\left(\frac{b_1}{b_2}\right)\cdot\frac{\partial v}{\partial z},\quad \left(\frac{u(x,y,z)\cdot\dfrac{\partial v}{\partial x}}{v(x,y,z)}\right)\cdot\left(\frac{v(x,y,z)\cdot\dfrac{\partial v}{\partial x}}{u(x,y,z)}\right)=-b_3\cdot a_1,\ \Rightarrow$$

$$\frac{\partial v}{\partial x}=\sqrt{(-b_3\cdot a_1)},\ \Rightarrow\ u(x,y,z)=-\left(\frac{b_3}{\sqrt{(-b_3\cdot a_1)}}\right)\cdot v(x,y,z),\quad \Rightarrow\ b_3=a_5,$$

$$\frac{\partial u}{\partial y}=-\left(\frac{b_4}{b_3}\right)\cdot\frac{\partial v}{\partial x},\ \Rightarrow\ \left(\frac{a_2}{a_1}\right)=\left(\frac{b_4}{b_3}\right),\quad b_5=-b_4,\quad \frac{\partial}{\partial y}(v(x,y,z))=-b_6,\quad b_7=-b_1, \tag{4.2}$$

$$\frac{\partial}{\partial y}p(x,y,z)=-b_8\cdot v(x,y,z),\quad \frac{\partial}{\partial x}p(x,y,z)=-a_8\cdot\left(\frac{b_3}{\sqrt{(-b_3\cdot a_1)}}\right)\cdot v(x,y,z),\quad f_y=-b_9\cdot v(x,y,z),$$

- and Eq. (3.7) yields

$$\left(\frac{u(x,y,z)\cdot\dfrac{\partial u}{\partial z}}{w(x,y,z)}\right)\cdot\left(\frac{w(x,y,z)\cdot\dfrac{\partial u}{\partial z}}{u(x,y,z)}\right)=-a_3\cdot c_1,\ \Rightarrow\ \frac{\partial u}{\partial z}=\sqrt{(-a_3\cdot c_1)},\ \Rightarrow\ \frac{\partial v}{\partial z}=-\frac{\sqrt{(a_3\cdot c_1\cdot b_3\cdot a_1)}}{b_3},$$

$$w(x,y,z)=-\left(\frac{a_3}{\sqrt{(-a_3\cdot c_1)}}\right)\cdot u(x,y,z)=\left(\frac{a_3}{\sqrt{(-a_3\cdot c_1)}}\right)\cdot\left(\frac{b_3}{\sqrt{(-b_3\cdot a_1)}}\right)\cdot v(x,y,z),\quad \frac{\partial w}{\partial x}=-\left(\frac{c_2}{c_1}\right)\cdot\frac{\partial u}{\partial z},\ \Rightarrow$$

$$\left(\frac{a_4}{a_3}\right)=\left(\frac{c_2}{c_1}\right),\ -\frac{v(x,y,z)\cdot\dfrac{\partial w}{\partial y}}{w(x,y,z)}=c_3,\ \Rightarrow\ \frac{\partial w}{\partial y}=\sqrt{(-b_1\cdot c_3)},\ \frac{\partial w}{\partial y}=-\left(\frac{c_3}{c_4}\right)\cdot\frac{\partial v}{\partial z},\ \Rightarrow\ \left(\frac{c_3}{c_4}\right)=\left(\frac{b_1}{b_2}\right),\ c_5=-c_1, \tag{4.3}$$

$$c_6=-c_4,\ \frac{\partial}{\partial z}(w(x,y,z))=-c_7,\ \frac{\partial}{\partial z}p(x,y,z)=-c_8\cdot w(x,y,z)=-c_8\cdot\left(\sqrt{\frac{a_3\cdot b_3}{a_1\cdot c_1}}\right)\cdot v(x,y,z),\ f_z=-c_9\cdot w(x,y,z).$$

So, the space part of the solution should be presented as below:

$$u = -\left(\frac{b_3}{\sqrt{(-b_3\cdot a_1)}}\right)\cdot v(x,y,z) = -b_3\cdot x + b_3\cdot b_6\cdot y + \sqrt{(-a_3\cdot c_1)}\cdot z$$

$$v = x - b_6\cdot y - \frac{\sqrt{(-a_3\cdot c_1)}}{b_3}\cdot z, \qquad (4.4)$$

$$w = \left(\sqrt{\frac{a_3\cdot b_3}{a_1\cdot c_1}}\right)\cdot v(x,y,z) = \sqrt{(-\frac{a_3}{c_1})}\cdot|b_3|\cdot x - b_6\cdot\left(\sqrt{\frac{a_3\cdot b_3}{a_1\cdot c_1}}\right)\cdot y - a_3\cdot z,$$

$$(a_1\cdot b_3) = -1,\quad a_2 = b_6,\quad |a_3|\cdot|b_3| = a_3\cdot b_3,\quad a_3 = c_7,\quad a_5 = b_3,\quad a_8\cdot b_3\cdot b_6 = b_8,\quad a_8 = \frac{c_8}{|a_1|\cdot|c_1|},$$

$$p(x,y,z) = a_8\cdot\left(-b_3\cdot b_6\cdot x\cdot y - \sqrt{(-a_3\cdot c_1)}\cdot x\cdot z + b_6\sqrt{(-a_3\cdot c_1)}\cdot y\cdot z + \frac{b_3}{2}x^2 + \left(\frac{b_3\cdot(b_6)^2}{2}\right)\cdot y^2 + \frac{|c_1|\cdot|a_3|}{2b_3}\cdot z^2\right)$$

Thus, if we choose for simplicity the proper constants as

$$c_1 = -1,\quad a_8 = 1\quad (\Rightarrow\ b_8 = b_3\cdot b_6,\ c_8 = 1/|b_3|\ ) \qquad (4.5)$$

- the space part of the solution should be presented as below ( $|a_3|\cdot|b_3| = a_3\cdot b_3$ ):

$$u = -b_3\cdot x + b_3\cdot b_6\cdot y + \sqrt{a_3}\cdot z$$

$$v = x - b_6\cdot y - \frac{\sqrt{a_3}}{b_3}\cdot z, \qquad (4.6)$$

$$w = \sqrt{a_3}\cdot|b_3|\cdot x - b_6\cdot|b_3|\cdot\sqrt{a_3}\cdot y - a_3\cdot z,$$

$$p(x,y,z) = -b_3\cdot b_6\cdot x\cdot y - \sqrt{a_3}\cdot x\cdot z + b_6\sqrt{a_3}\cdot y\cdot z + \frac{b_3}{2}x^2 + \left(\frac{b_3\cdot(b_6)^2}{2}\right)\cdot y^2 + \frac{|a_3|}{2b_3}\cdot z^2$$

## 5. Time-dependent part of exact solution.

As for the structure of time-dependent part of exact solution (3.1) with space part (4.6), it could be obtained from system of Eqs. (3.4) which should be transformed as below:

$$\begin{cases} \dfrac{dU(t)}{dt} = V(t)\cdot U(t)\cdot a_2 + W(t)\cdot U(t)\cdot a_3 + U(t)^2\cdot a_5 + P(t)\cdot a_8 + a_9\,, \\ \dfrac{dV(t)}{dt} = W(t)\cdot V(t)\cdot b_2 + U(t)\cdot V(t)\cdot b_3 + V^2(t)\cdot b_6 + P(t)\cdot b_8 + b_9\,, \\ \dfrac{dW(t)}{dt} = U(t)\cdot W(t)\cdot c_2 + V(t)\cdot W(t)\cdot c_3 + W^2(t)\cdot c_7 + P(t)\cdot c_8 + c_9\,, \end{cases} \tag{5.1}$$

- where $a_2 = b_6$, $|a_3|\cdot|b_3| = a_3\cdot b_3$, $a_5 = b_3$, $a_8 = 1$, $b_8 = b_3\cdot b_6$, $c_7 = a_3$, $c_8 = 1/|b_3|$; besides, it should be accomplished along with the continuity equation (3.3):

$$U(t)\frac{\partial u}{\partial x} + V(t)\frac{\partial v}{\partial y} + W(t)\frac{\partial w}{\partial z} = 0\,, \;\Rightarrow\; U(t)\cdot b_3 + V(t)\cdot b_6 + W(t)\cdot a_3 = 0\,,$$

- so, we have 4 equations for the obtaining of 4 functions $U(t)$, $V(t)$, $W(t)$, $P(t)$.

Along with the invariant from the continuity equation (besides, $a_9\cdot b_3 + b_9\cdot b_6 + c_9\cdot a_3 = 0$)

$$\frac{d\left( U(t)\cdot b_3 + V(t)\cdot b_6 + W(t)\cdot a_3 \right)}{dt} = 0\,,$$

- a system of Eqs. (5.1) immediately yields the invariant for function $P(t)$ as below

$$\begin{aligned} P(t)\cdot\left(a_8\cdot b_3 + b_8\cdot b_6 + c_8\cdot a_3\right) &= -\,[V(t)\cdot U(t)\cdot a_2 + W(t)\cdot U(t)\cdot a_3 + U(t)^2\cdot a_5\,]\cdot b_3\; - \\ &-[W(t)\cdot V(t)\cdot b_2 + U(t)\cdot V(t)\cdot b_3 + V^2(t)\cdot b_6]\cdot b_6 - [U(t)\cdot W(t)\cdot c_2 + V(t)\cdot W(t)\cdot c_3 + W^2(t)\cdot c_7]\cdot a_3\,, \end{aligned} \tag{5.2}$$

- thus, we should exclude expression (5.2) for function $P(t)$ from the analysis of equations of system (5.1) and also we should exclude the continuity equation, which means that one of 3 functions $U(t)$, $V(t)$, $W(t)$ is the linear combination of two others:

$$W(t) \; = \; - \; U(t)\cdot\frac{b_3}{a_3} \; - \; V(t)\cdot\frac{b_6}{a_3} \tag{5.3}$$

So, analyzing the system (5.1), we finally should obtain the system of 2 ordinary differential equations of the 1-st order for any 2 of 3 functions $U(t)$, $V(t)$, $W(t)$ (*the last 3-rd function could be obtained by expressing it from the continuity equation above*).

These governing ODE-equations form together a system of 2 *Riccati*-type equations in regard to each other, which is the system of 2 ordinary differential equations of the 1-st kind with the right parts, consisting of polynomials of the 2-nd extent in regard to the functions $U(t)$, $V(t)$, $W(t)$.

*Riccati* type of equations has no analytical solution in general case [6]. We should note also that a modern methods exist for obtaining of the solution of Riccati equations with a good approximation [7-9]. But if we choose a proper constants for the system (5.1):

$$a_9 = b_9 = c_9 = 0, \qquad (a_8{\cdot}b_3 + b_8{\cdot}b_6 + c_8{\cdot}a_3) = 1, \; \Rightarrow$$

$$(b_3 + b_3{\cdot}(b_6)^2 + a_3/|b_3|) = 1, \;\; \Rightarrow \;\; a_3 = |b_3| \; - b_3 \cdot|b_3| - b_3{\cdot}|b_3|{\cdot}(b_6)^2, \tag{5.4}$$

- the 1-st and 2-nd equations of system (5.1) could be transformed as presented below:

$$\begin{cases} \dfrac{dU(t)}{dt} = V(t)\cdot U(t)\cdot\left(b_3\cdot\left(\dfrac{b_2\cdot b_6}{a_3}+c_3-2b_6\right)+b_6\cdot(c_2-b_3)\right) + \\ \qquad + \; U(t)^2\cdot b_3\cdot(c_2-b_3) \; + \; V^2(t)\cdot b_6\cdot\left(\dfrac{b_2\cdot b_6}{a_3}+c_3-2b_6\right) \; , \\ \\ \dfrac{dV(t)}{dt} = \; U(t)\cdot V(t)\cdot\left(b_3\cdot\left(1-\dfrac{b_2}{a_3}\right)+b_6\cdot b_8\cdot(c_2-b_3)+b_3\cdot b_8\cdot\left(\dfrac{b_2\cdot b_6}{a_3}+c_3-2\,b_6\right)\right) + \\ \qquad + \; V^2(t)\cdot\left(b_6\cdot\left(1-\dfrac{b_2}{a_3}\right)+b_6\cdot b_8\cdot\left(\dfrac{b_2\cdot b_6}{a_3}+c_3-2b_6\right)\right) \; + \; U^2(t)\cdot b_3\cdot b_8\cdot(c_2-b_3) \; , \end{cases} \tag{5.5}$$

- where $a_2 = b_6$, see (4.4). Besides, if we additionally choose $b_2 = a_3$, system (5.5) above could be reduced to the simplified regular form below:

$$\begin{cases} \dfrac{dU(t)}{dt} = \left( b_8 \cdot U(t) + C \right) \cdot U(t) \cdot \left( b_3 \cdot (c_3 - b_6) + b_6 \cdot (c_2 - b_3) \right) + \\ \qquad + U(t)^2 \cdot b_3 \cdot (c_2 - b_3) + \left( b_8 \cdot U(t) + C \right)^2 \cdot b_6 \cdot (c_3 - b_6) \ , \\ \dfrac{dV(t)}{dt} = b_8 \cdot \dfrac{dU(t)}{dt} \quad \Rightarrow \quad V(t) = b_8 \cdot U(t) + C, \quad C = const, \end{cases} \tag{5.6}$$

- where the 1-st equation of system (5.6) has a proper solution below ($b_8 = b_3 \cdot b_6$):

$$\frac{dU(t)}{(A \cdot U^2(t) + B \cdot U(t) + D)} = dt \, , \tag{5.7}$$

$$A = \left( b_8 \cdot b_3 \cdot (c_3 - b_6) + b_8 \cdot b_6 \cdot (c_2 - b_3) + b_3 \cdot (c_2 - b_3) + (b_8)^2 \cdot b_6 \cdot (c_3 - b_6) \right)$$

$$B = C \cdot \left( b_3 \cdot (c_3 - b_6) + b_6 \cdot (c_2 - b_3) + 2 b_8 \cdot b_6 \cdot (c_3 - b_6) \right), \quad D = C^2 \cdot b_6 \cdot (c_3 - b_6)$$

The left side of expression (5.7) could be transformed to the proper *elliptical* integral [10] in regard to function $U(t)$:

$$\int \frac{dU(t)}{(A \cdot U^2(t) + B \cdot U(t) + D)} = \begin{cases} \dfrac{2}{\sqrt{\Delta}} \arctan\left( \dfrac{2A \cdot U(t) + B}{\sqrt{\Delta}} \right), & \Delta > 0 \\ -\dfrac{2}{\sqrt{-\Delta}} Arth\left( \dfrac{2A \cdot U(t) + B}{\sqrt{-\Delta}} \right), & \Delta < 0 \end{cases} \qquad \Delta = (4A \cdot D - B^2) \tag{5.8}$$

## 6. <u>Discussion.</u>

In fluid mechanics, a lot of authors have been executing their researches to obtain the analytical solutions of Euler and Navier-Stokes equations [11], even for 3D case of *compressible* gas flow [12]. But there is an essential deficiency of non-stationary solutions indeed.

Our presentation (3.1) of the non-stationary solutions of 3D Euler equations (1.1)-(1.2) for incompressible flow is considered here. The spatial part of such a solution is determined by equalities (4.4), under the given initial conditions; but the time-dependent part is determined by Eqs. (5.1)-(5.3).

Besides, the real example of exact solution is obtained. The spatial part of such a solution is presented by the equalities (4.5)-(4.6), but the time-dependent part is presented by equalities (5.2)-(5.3), (5.4)-(5.8).

Also, we should especially note that the components of flow velocity (4.6) of the solution (3.1) will be uniformly increasing when $(x, y, z) \to \infty$. So, such a solution should be defined within the limited domain of the meanings of variables $(x, y, z)$, it should be given by the initial conditions.

The explicit solutions for $U(t)$ can easily be obtained from Eqs. (5.7) and (5.6), therefore, the explicit form of the special solutions (3.1) should be provided:

$$u_1 = U(t)\cdot u(x,y,z),\ \ u_2 = V(t)\cdot v(x,y,z),\ \ u_3 = W(t)\cdot w(x,y,z),\ \ p = P(t)\cdot p(x,y,z),$$

- where

$$u = -b_3\cdot x + b_3\cdot b_6\cdot y + \sqrt{a_3}\cdot z$$

$$v = x - b_6\cdot y - \frac{\sqrt{a_3}}{b_3}\cdot z,$$

$$w = \sqrt{a_3}\cdot|b_3|\cdot x - b_6\cdot|b_3|\cdot\sqrt{a_3}\cdot y - a_3\cdot z,$$

$$p(x,y,z) = -b_3\cdot b_6\cdot x\cdot y - \sqrt{a_3}\cdot x\cdot z + b_6\sqrt{a_3}\cdot y\cdot z + \frac{b_3}{2}x^2 + \left(\frac{b_3\cdot(b_6)^2}{2}\right)\cdot y^2 + \frac{|a_3|}{2b_3}\cdot z^2,$$

$$P(t) = -[V(t)\cdot U(t)\cdot b_6 + W(t)\cdot U(t)\cdot a_3 + U(t)^2\cdot b_3]\cdot b_3 -$$

$$-[W(t)\cdot V(t)\cdot a_3 + U(t)\cdot V(t)\cdot b_3 + V^2(t)\cdot b_6]\cdot b_6 - [U(t)\cdot W(t)\cdot c_2 + V(t)\cdot W(t)\cdot c_3 + W^2(t)\cdot a_3]\cdot a_3\ ,$$

$$W(t) = -U(t)\cdot\frac{b_3}{a_3} - V(t)\cdot\frac{b_6}{a_3}, \qquad V(t) = (b_3\cdot b_6)\cdot U(t) + C, \quad C = const \ ,$$

- here we should choose $a_3 = \{|b_3| - b_3 \cdot|b_3| - b_3\cdot|b_3|\cdot(b_6)^2\} \neq 0, \rightarrow b_3 \cdot(1+ (b_6)^2) \neq 1$, but the key function $U(t)$ should be given as below ($b_8 = b_3\cdot b_6$):

$$\begin{cases} U(t) = \dfrac{\sqrt{\Delta}\cdot\tan\left(\left(\dfrac{\sqrt{\Delta}}{2}\right)\cdot t\right) - B}{2A}, & \Delta > 0 \\ U(t) = \dfrac{\sqrt{-\Delta}\cdot\tanh\left(-\left(\dfrac{\sqrt{-\Delta}}{2}\right)\cdot t\right) - B}{2A}, & \Delta < 0 \end{cases} \qquad \Delta = (4A\cdot D - B^2)$$

$$A = \left(b_8\cdot b_3\cdot(c_3 - b_6) + b_8\cdot b_6\cdot(c_2 - b_3) + b_3\cdot(c_2 - b_3) + (b_8)^2\cdot b_6\cdot(c_3 - b_6)\right)$$

$$B = C\cdot\left(b_3\cdot(c_3 - b_6) + b_6\cdot(c_2 - b_3) + 2b_8\cdot b_6\cdot(c_3 - b_6)\right), \quad D = C^2\cdot b_6\cdot(c_3 - b_6)$$

For example, if we choose $c_3 = b_6$ ($C \neq 0$) it should simplify the expression for $U(t)$:

$$U(t) = \frac{\sqrt{-\Delta}\cdot\tanh\left(-\left(\dfrac{\sqrt{-\Delta}}{2}\right)\cdot t\right) - B}{2A}, \qquad \Delta = -B^2$$

$$A = b_3\cdot(c_2 - b_3)\cdot\left(1 + (b_6)^2\right), \quad B = C\cdot b_6\cdot(c_2 - b_3), \quad D = 0$$

- and, if we additionally choose $c_2 = 2b_3$, $b_3 = 2/C$ ($C \neq 0$), $b_6 = 1$, we should obtain (see Fig.1-5)

$$U(t) = \frac{C^2}{8}\cdot\left(\tanh(-t) - 1\right), \qquad \Delta = -4, \quad A = (b_3)^2\cdot 2, \quad B = 2, \quad D = 0$$

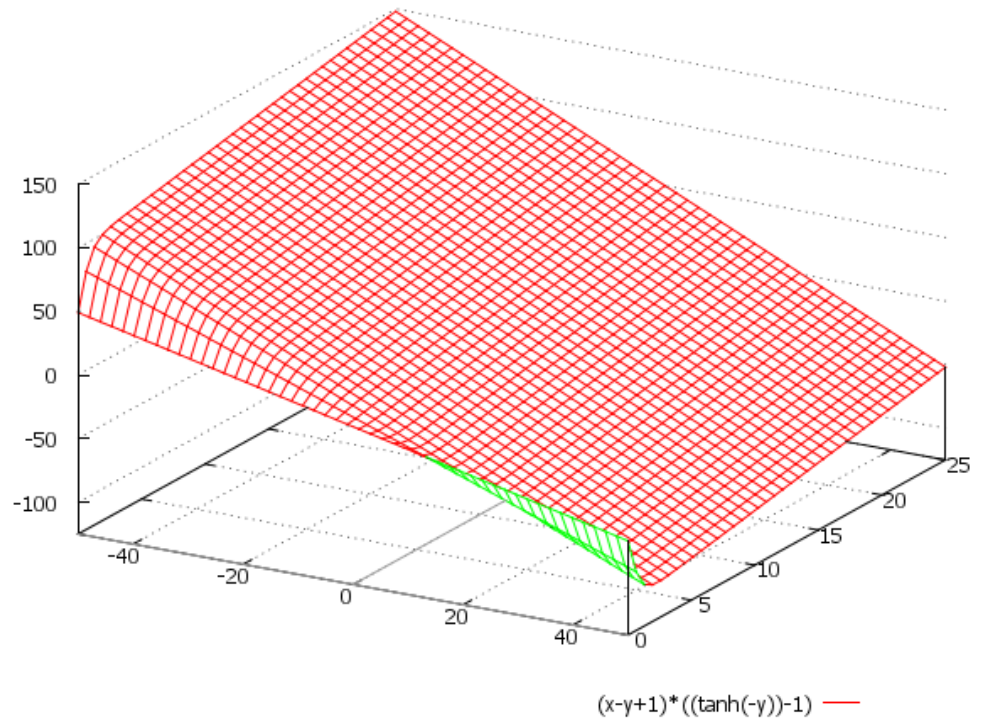

Fig.1. A *schematic* plot of the function ~ ($x$-$y$+1)*{tanh(-$t$) - 1},

here we designate: $x \in (-50, 50)$, $t = y \in (0, 25)$.

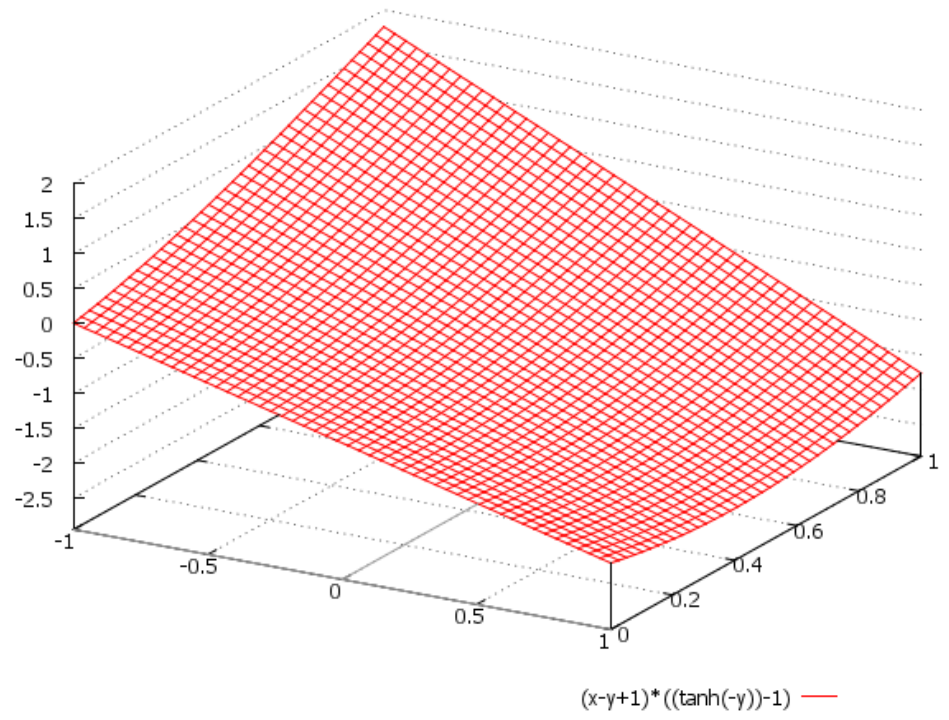

Fig.2. A *schematic* plot of the function ~ ($x$-$y$+1)*{tanh(-$t$) - 1},

here we designate: $x \in (-1, 1)$, $t = y \in (0, 1)$.

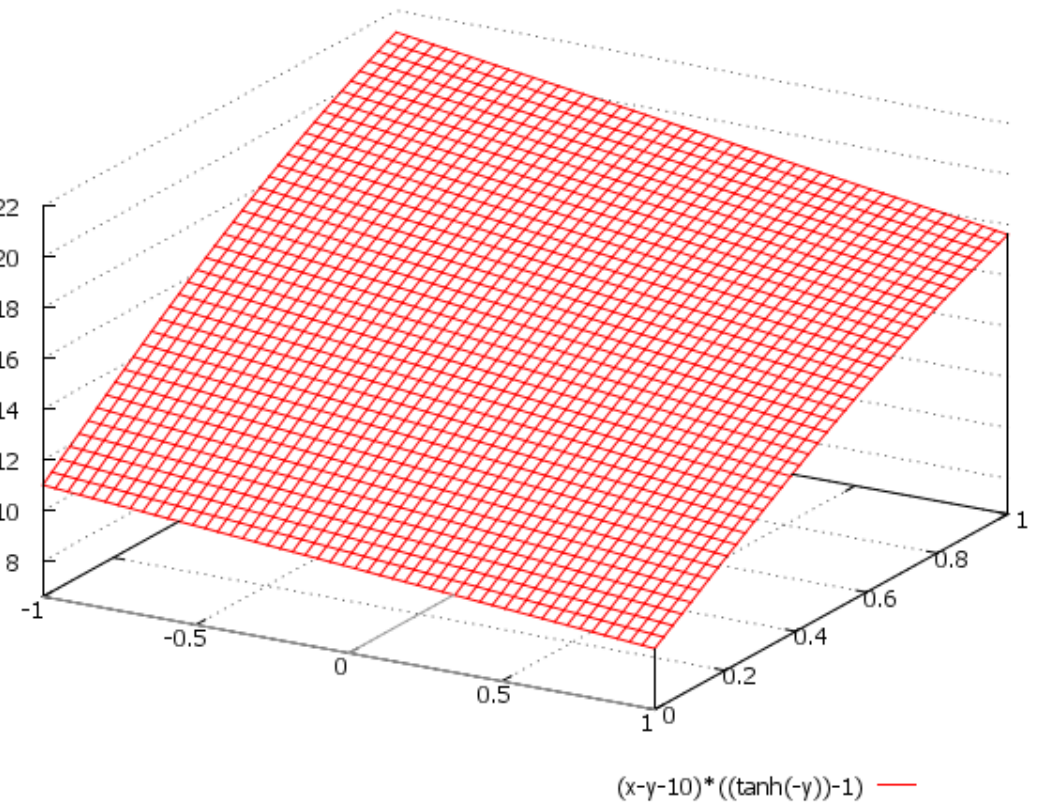

Fig.3. A *schematic* plot of the function ~ ($x$-$y$-10)*{tanh(-$t$) - 1},

here we designate: $x \in (-1, 1)$, $t = y \in (0, 1)$.

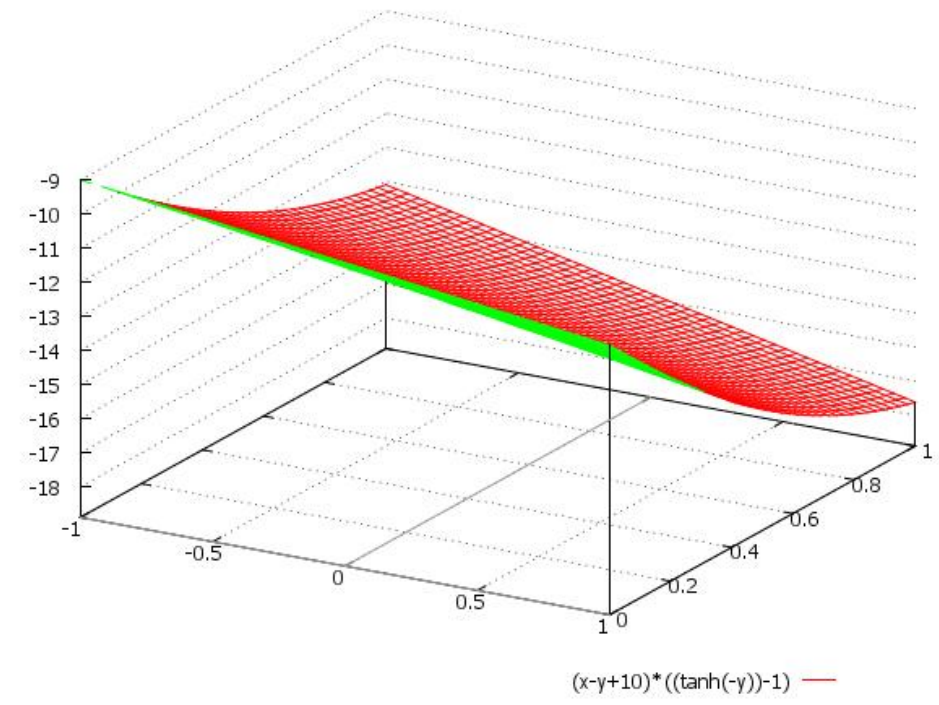

Fig.4. A *schematic* plot of the function ~ (*x*-*y*+10)*{tanh(-*t*) - 1},

here we designate: $x \in (-1, 1)$, $t = y \in (0, 1)$.

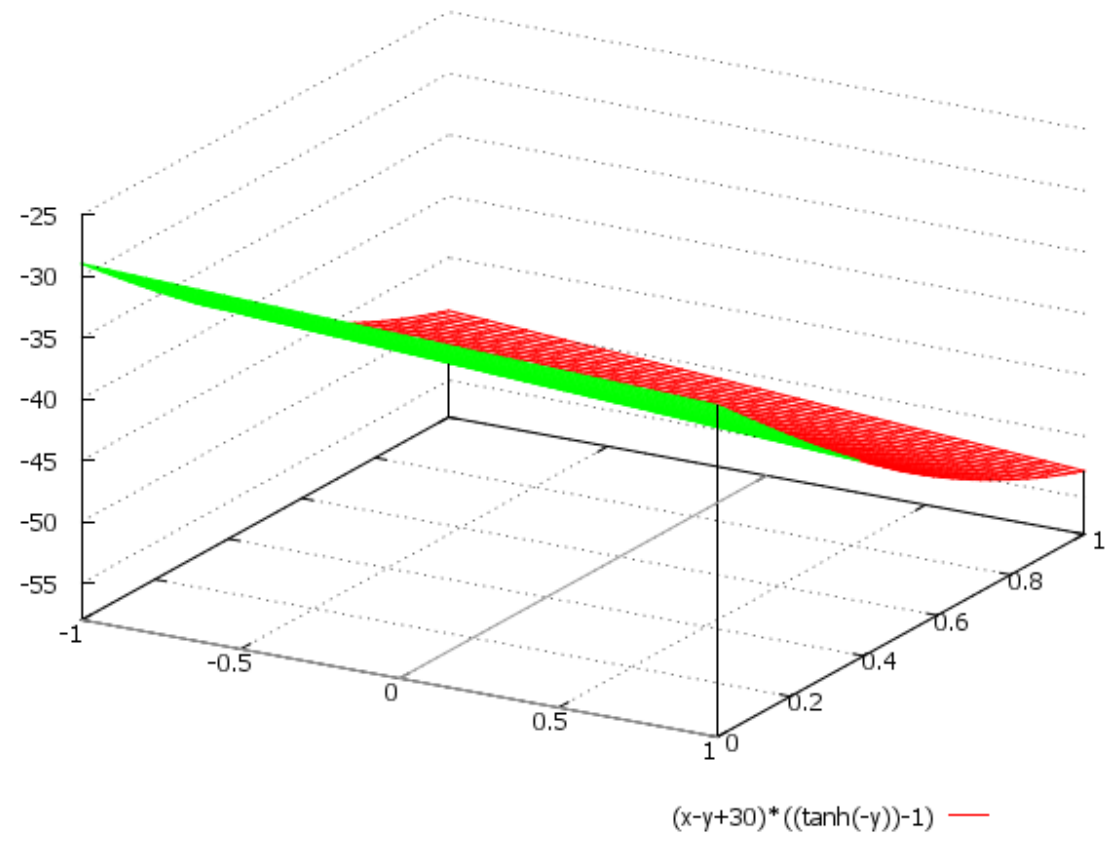

Fig.5. A *schematic* plot of the function ~ (*x*-*y*+30)*{tanh(-*t*) - 1},

here we designate: $x \in (-1, 1)$, $t = y \in (0, 1)$.

We assume at Fig.1-5 that in the expressions (*x*-*y*-10), (*x*-*y*+1), (*x*-*y*+10), (*x*-*y*+30) set of meanings {-10, 1, 10, 30} is varying according to the varying of the range of variable *z*; besides, the factor {tanh(-*t*) - 1} could be schematically presented (for imagination of the plots of solutions) by the changing of parameter *t* to variable *y*, for example.

At Fig. 6-8 we schematically imagined solutions for the case $\Delta>0$:

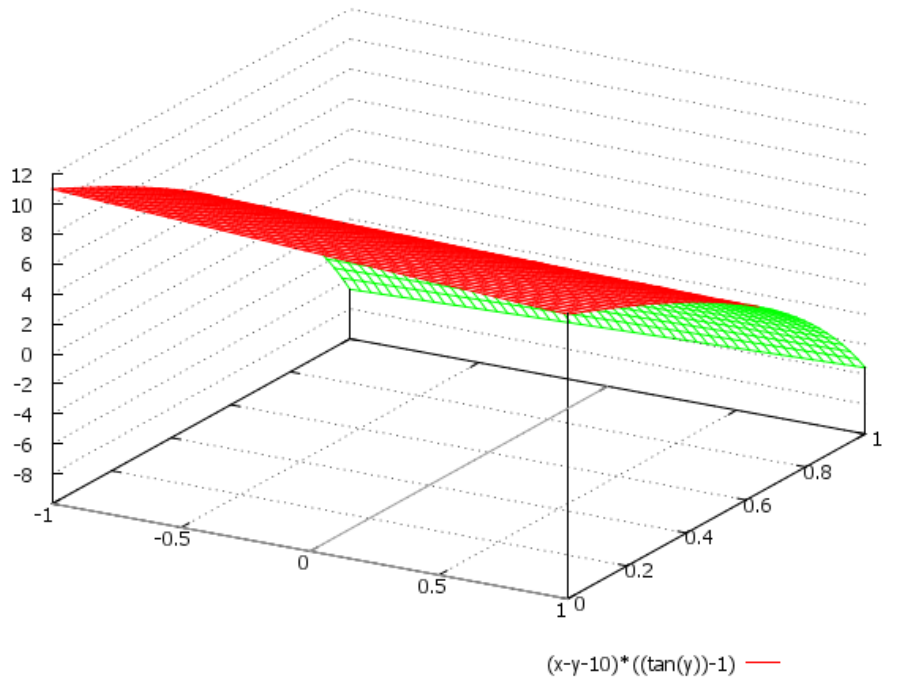

Fig.6. A *schematic* plot of the function ~ ($x$-$y$-10)*{tan($t$) - 1},

here we designate: $x \in (-1, 1)$, $t = y \in (0, 1)$.

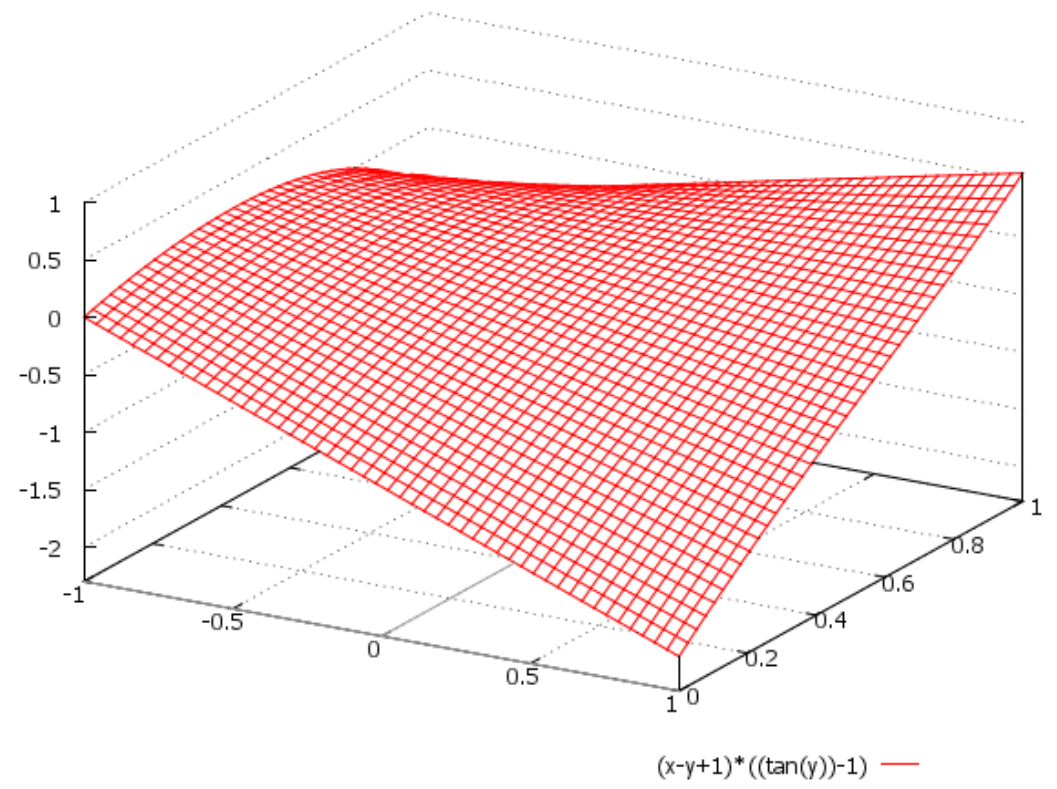

Fig.7. A *schematic* plot of the function ~ ($x$-$y$+1)*{tan($t$) - 1},

here we designate: $x \in (-1, 1)$, $t = y \in (0, 1)$.

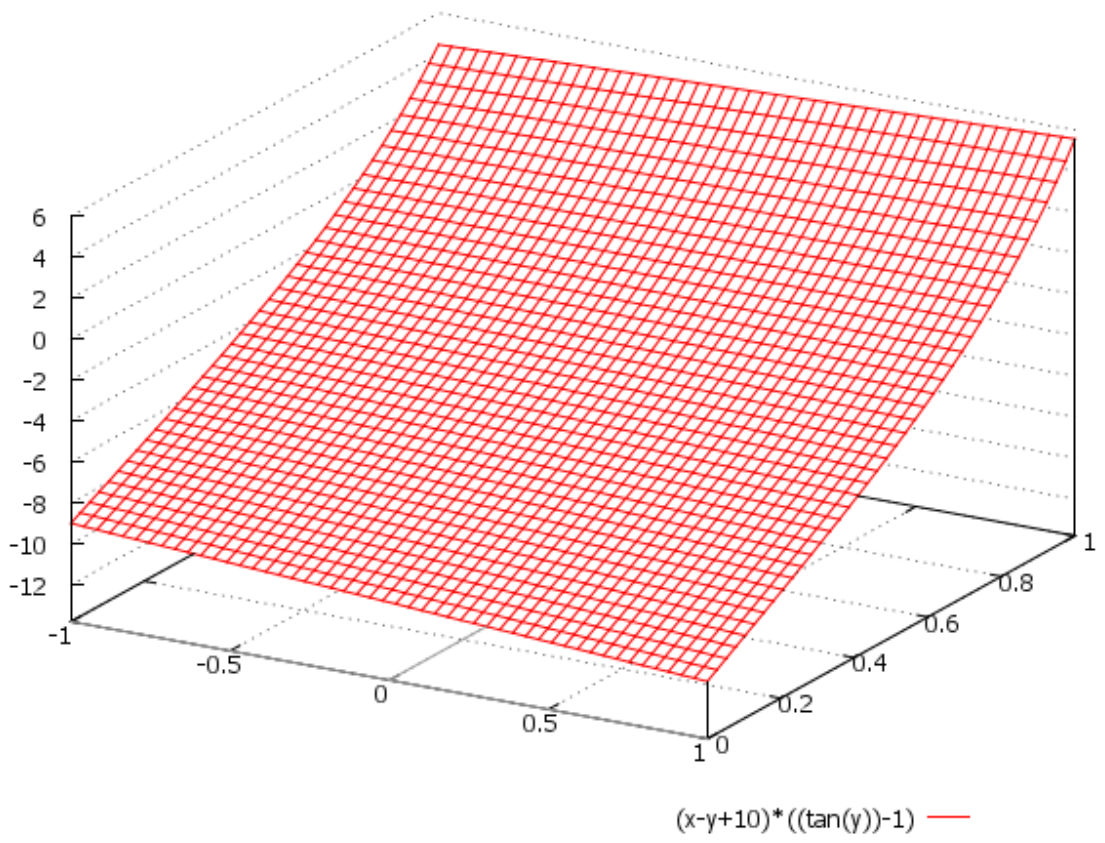

Fig.8. A *schematic* plot of the function ~ ($x$-$y$+10)*{tan($t$) - 1},

here we designate: $x \in (-1, 1)$, $t = y \in (0, 1)$.

Also, we should note that since some solutions are unbounded (see for instance Eq. (5.8) for $\Delta>0$), such a solution should be defined within the limited range of the meanings of time-parameter $t$ (it should be given by the initial conditions).

Besides, we should additionally note that the only periodic (and unbounded) solutions are the ones given by $U(t)$ for $\Delta > 0$, since the hyperbolic tangent is a non periodic but bounded function.

## 7. <u>Conclusion.</u>

A new presentation of non-stationary solutions of 3D Euler equations for incompressible inviscid flow is considered here. Such a solution is the product of 2 separated parts: - spatial and the time-dependent parts.

Spatial part of a solution could be determined if we substitute such a solution to the equations of motion (equation of momentum) under the demand of scale-similarity in regard to the proper component of spatial velocity. So, the time-dependent part of equations of momentum should depend on the time-parameter only.

The main result, which should be outlined, is that the governing (time-dependent) ODE-system consist of 2 *Riccati*-type equations in regard to each other, which has no solution in general case. But we obtain conditions when each component of time-dependent part is proved to be determined by the proper *elliptical* integral in regard to the time-parameter $t$, which is a generalization of the class of inverse periodic functions. Thus, by the proper obtaining of re-inverse dependence of a solution from time-parameter we could present the expression for each component of motion as a set of periodic cycles.